\documentclass[prb]{revtex4}

\usepackage{graphicx}
\usepackage{dcolumn}
\usepackage{bm}
\usepackage{bbm}
\usepackage{amsmath}

\begin{document}

\title{Effective interactions in multi-band systems from constrained summations}
\author{Carsten Honerkamp}
\affiliation{Institute for Theoretical Solid Physics, RWTH Aachen University \\ and JARA - FIT Fundamentals of Future Information Technology}
\date{December 21, 2011}

\begin{abstract}
In correlated electron materials, the application of many-body techniques for the study of interaction effects or unconventional superconductivity often requires the formulation of an effective low-energy model that contains only the relevant bands near the Fermi level. However the bands away from the Fermi level are known to renormalize the low-energy interactions substantially. Here we compare different schemes to derive low-energy effective theories for interacting electrons in solids. The frequently used constrained random phase approximation (cRPA) is identified as a particular resummation of  higher-order interaction terms that includes important virtual corrections. We then propose an adapted functional renormalization group (fRG) scheme that includes the cPRA, but also allows one to go beyond the cRPA approximation. We study a simple two-band model in order to explore the differences between the approximations.
\end{abstract}

\maketitle
\section{Introduction}
Due to the complexity of many condensed-matter systems, the derivation of effective low-energy theories has always represented a core problem of many-particle physics. Often, these low-energy models are still complicated enough to provide challenges for a large community of researchers. For example in the field of strongly correlated electrons, understanding the properties of Hubbard, Anderson-impurity, or Kondo-type models has been a central topic of the recent decades, which has in parts been pursued with very little material-specific input, and without the necessity or also possibility to derive the model parameters in theory with higher precision. 
However, with a larger toolbox of more quantitative many-body techniques for solving the low-energy problem available, a more precise determination (if at all possible) of effective low-energy models in correlated electron systems becomes increasingly important. For example, the recent theoretical research in the iron arsenides makes clear that details of the effective model are relevant at least for the gap structure of the leading superconducting instabilities\cite{kurokipnictides,maier,platt}. Furthermore, in the field of graphene physics, the occurrence of interesting many-body states might depend on whether the interaction parameters in a description reduced to the two lowest $\pi$-bands exceed certain threshold values\cite{wehling,honerkamp_honey,raghu,meng}. 

The constrained random phase approximation (cRPA)\cite{cRPA,imadamiyakereview} has become a standard tool for deriving interactions in a localized basis for such effective low-energy few-band models for solids. Compared to the straightforward Coulomb matrix elements between the localized Wannier states  forming the low-energy  band, which would be the simplest choice as effective interaction parameters, in the cRPA interband transitions involving particle-hole pairs with at least one intermediate particle outside the low-energy range are included as well. These particle-hole diagrams are then summed up to infinite order, which usually leads to a significant reduction of the local Hubbard interaction parameters. In the cRPA procedure, the particle-hole screening with two intermediate particles in the low-energy sector is excluded, as these processes should be dealt with in the solution of the resulting low-energy problem. In general, the cRPA procedure generates an interaction that depends on the frequency transfer through the diagrams\cite{cRPAom}. While this can have some effect on excited states, it is found that the low-frequency dependence in rather weak such that working with the low-frequency limit can be considered a good approximation\cite{imadamiyakereview}. Furthermore, in contrast with the constrained local-density-approximation\cite{cLDA} (cLDA) also used in the field, the cRPA in general produces orbitally dependent interaction parameters. This gives an extra handle to understand different behaviors in complex multi-orbital systems.

The cRPA has been accepted well and is being applied with appealing results to a growing number of situations (see, e.g., Refs.~\onlinecite{wehling,miyake,sasioglu}). Theoretically, there are however various questions that should be considered in order to understand better the potential of this approach. First of all, the random phase approximation can be argued to capture physically relevant contributions, but a priori there seems to be no argument why this approximation really captures all important physics. For the screening of long-range part of $1/q^2$  Coulomb interactions ($q$ denotes the wavevector-transfer), it might be sufficient to focus on the RPA series, as for any vertex correction one would integrate over a factor $1/q^2$ at a vertex, and hence, for any fixed order $n$ in the bare interactions, the RPA-type bubble sum of order $n$ is the leading term larger by a factor $\sim \left( \frac{k_F}{q}\right)^2$\cite{negele}. Note however that the main field of application of the cRPA is determining the short-range 'Hubbard' part of the interactions, and hence the $q \to 0$-part will most likely not give the full answer. For general $q$, if we ask why other one-loop terms are ignored, there appears to be no expansion parameter other than the inverse energy separation between high- and low-energy bands, but this enters in all one-loop diagrams in a similar way. Likewise, large-$N$ arguments do not seem applicable, as the Coulomb interaction does not imply any band-conservation rules.
Hence, a central point of this work is present a broader framework that contains more corrections than cRPA, that allows us to assess the validity of the cRPA scheme for a given model. 

A common strategy for deriving low-energy models are Wilsonian renormalization group (RG) methods\cite{wilson,wegner}. Here the degrees of freedom above a certain energy scale $s$ are integrated out, leading to an effective action for the remaining low-energy modes. With regard to the question which diagrams should be kept and which can be ignored, RG techniques for many-fermion systems\cite{solyom,shankar,salmhoferbook,metznerreview,kopietz} have been proven very useful, as they are capable of including all one-loop diagrams and hence go beyond standard single-channel summation like the RPA. Therefore restricted summations can be tested against a more complete picture that treats all one-loop diagrams on the same footing. Hence, the obvious strategy in this work is to explore the use of RG techniques in order to embed the cRPA in this broader context. We will see that the cRPA is already a clever extension of the standard truncation that is usually taken in the RG approach. These ideas will then allow us to propose improved derivation schemes of low-energy effective interactions.

This paper is organized as follows. In Sec. II we briefly reproduce the main equations of the cRPA formalism. Then, in Sec. III A we contrast this with the standard RG approach for fermionic low-energy effective theories. In Sec. III B we identify the cRPA as a particular resummation of terms that would have been neglected on the standard truncation level of the RG approach. 
In Sec. III C we then relate this resummation to Wick-ordered correlation functions. In Sec. III D we propose a functional RG (fRG) scheme that includes the cRPA but also includes many other terms of comparable importance. In Sec. IV we present a numerical study of a simple model system that allows us to monitor the differences between the approximation levels. Finally we conclude in Sec. V with a brief discussion. 

\section{Constrained random phase approximation}
Here we review the main construction principle of the constrained random phase approximation (cRPA). For more information, the reader is referred to the original papers by Aryasetiawan et al.\cite{cRPA} or a recent review by Imada and Miyake\cite{imadamiyakereview}. 

The standard set-up we consider here is a multiband system with 'low-energy' or 'target' bands near the Fermi level and 'high-energy' bands away from the Fermi level. We assume that the bands have been determined by an ab-initio technique such as density functional theory with local density approximation (DFT-LDA). Let $s$ be an energy scale that separates the two band complexes, target and high-energy bands. The situation of overlapping bands can also dealt with without major changes, but for simplicity we will not dwell on this issue here.

If interaction effects in the low-energy bands are important, it will be advantageous to derive a real-space few-orbital model for these bands, typically employing (maximally) localized Wannier functions\cite{vanderbilt,imadamiyakereview} formed with the low-energy bands as basis functions. The hopping dispersion between these orbitals is then obtained from the DFT bands. A first estimate for the interaction parameters can be obtained by computing Coulomb and exchange integrals between these Wannier states. Note that the values obtained this way depend on the target energy window.
A smaller target window usually corresponds to less localized Wannier orbitals, and will result in smaller interaction parameters (for comparisons between pure $d$-orbital models and more extended $dp$-models in the iron pnictides see, e.g., Refs. \onlinecite{miyake,anisimov}). It is however clear that this change arises just from the two different Wannier bases with a different charge spread of the basis functions localized on a specific site in the two representations. More importantly, the kernels $\frac{e^2}{|\vec{r}-\vec{r}'|}$ of the Coulomb or exchange integrals lack the effects of virtual excitations into the higher bands outside the target window. These processes induce, besides another changes to be discussed below, the screening that lead to a renormalized Coulomb interaction  $\frac{e^2}{\epsilon (\vec{r},\vec{r}',\omega) |\vec{r}-\vec{r}'|}$ with the dielectric constant $\epsilon (\vec{r},\vec{r}',\omega)$ of the fictitious insulator with the bands near the Fermi level cut out. This screening is responsible for a potentially significant reduction of the effective interaction parameters.

The main idea of the constrained RPA is now to include these virtual processes on the level of the random phase approximation ({\em assuming}Ê that vertex corrections are not overly important), by renormalizing the bare Coulomb interaction $v(\vec{q})$ to \begin{equation}
\label{vcrpa}
V^{cRPA} (q) = \frac{v(\vec{q}) }{ 1- P_s (q) v(\vec{q}) } \, .
\end{equation}
In this expression, $q=(\vec{q}, iq_0)$ combines the transferred wavevector $\vec{q}$ and Matsubara frequency $iq_0$, and the polarization particle-hole bubble $P_s (q) $ only contains contributions with at least one internal line in the the high-energy bands. Using the high-energy electron propagator $G_{>s}(k)$, with $k=(\vec{k},ik_0)$ for the fermionic quantum numbers (excluding spin dependences and not writing possible band indices), and the low-energy propagator $G_{<s}(k)$, is it is given by
\begin{equation}
\label{Ps}
P_s(q) = 2 \sum_k \left[ G_{>s}(k) G_{>s}(k+q) + G_{>s}(k) G_{<s}(k+q) + G_{<s}(k) G_{>s}(k+q) \right] \, . 
\end{equation}  
The factor of two in from of the wavevector- and Matsubara-sum is from the spin sum, assuming spin-rotational invariance. The first term in the square bracket is the high-energy polarization involving only bands away from the Fermi level, while the second and the third term correspond to 'mixed' diagrams with one propagator line away from the Fermi level, and the other within the low-energy bands. If we consider a simplified setting with just one low-energy band near the Fermi level and only one high-energy band, these contributions correspond to interband particle-hole pairs. From the construction it is clear that $P_s(q)$ should not lead to any divergences, as all energy denominators contributing at $T\to 0$ remain finite and at least of order $s$, i.e. the energy separation between Fermi level and the high-energy bands. However, one should expect a wavevector- and frequency-dependence playing a role at $q_0 \sim s$\cite{imadamiyakereview}. 
\begin{figure}
 \begin{center}
 \includegraphics[width=.75\linewidth]{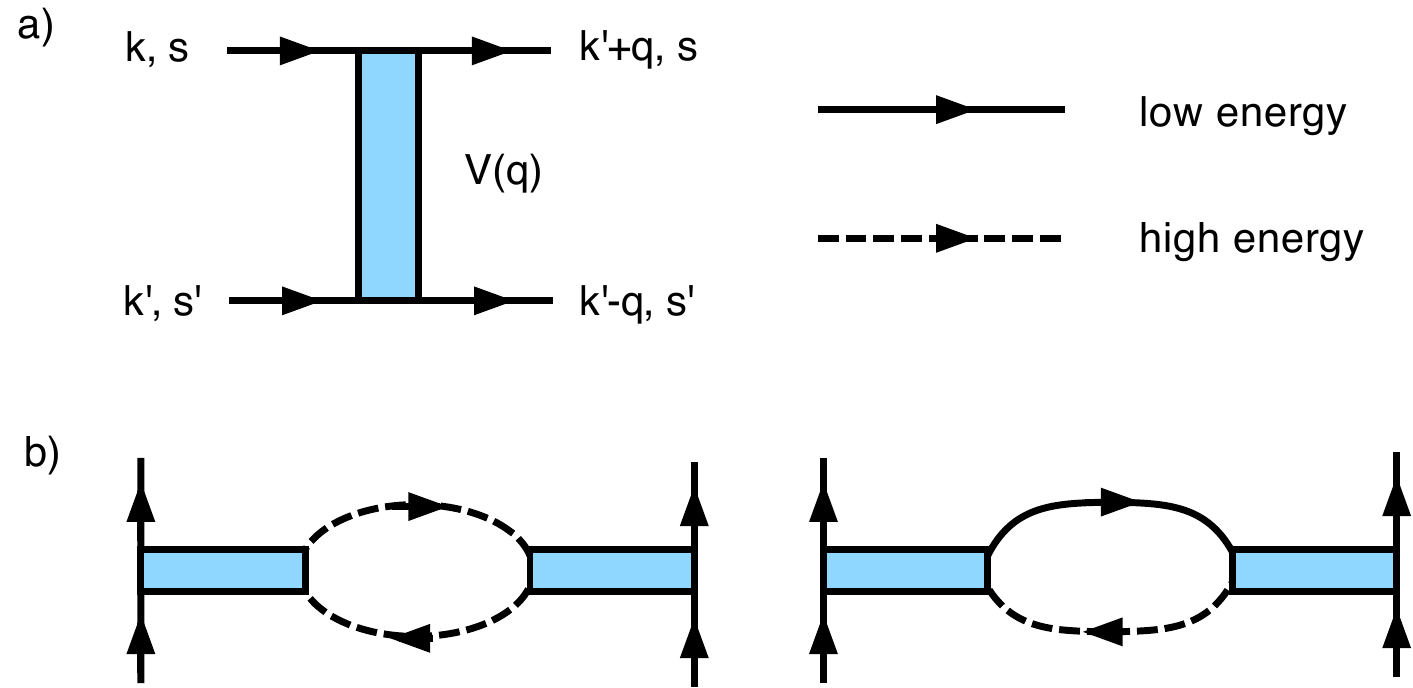}
 \end{center}
 \caption{a) Left side: Diagrammatic expression for the bare interaction $v(\vec{q})$ with the incoming and outgoing quantum numbers for wavevector and Matsubara frequency $k=(\vec{k},ik_0)$ and spin $s$. The spin projection is conserved along the short edge of the box.  Right side: The solid lines denote low-energy propagators, the dashed lines high-energy propagators. b) Two types of second-order corrections, summed together to infinite order in the cRPA. The term on the left side has two internal high-energy propagators, while on the right side, in the 'mixed' diagram, one internal line is in the low-energy sector.}
\label{IntBubbles}
\end{figure}
In our notation, for $iq_0 \to 0$, the partial poalization $P_s(q)$ becomes real and negative, and therefore the effective interaction $V^{cRPA} (\vec{q},0)$ is reduced compared to the bare value $v(\vec{q})$. As the mixed terms $G_{>s}(k) G_{<s}(k+q) + G_{<s}(k) G_{>s}(k+q)$ in Eq. \ref{Ps} with one line in the target bands have smaller energy denominators, the main contributions should come from the mixed diagrams. In realistic many-band calculations, the reduction factor can be $0.4 $ to $0.5$ for the $\pi$-bands of graphene or graphite\cite{wehling} or even $0.17$ to $0.3$ for various low-energy models for iron arsenide superconductors\cite{miyake}. Of course such strong renormalizations and their dependence on the material parameters are important input information for  any modeling of correlation effects beyond the DFT.

\section{Embedding the cRPA into the renormalization group}
\subsection{Renormalization group approach to effective theories}
Here, we outline the renormalization group (RG) approach to effective theories, in order to see how the cRPA can be related to this. Focussing on the fermionic case, let us again consider an energy scale $s$ defining a certain distance to the Fermi level, and let us assume that $s$ is located in band gaps between bands near the Fermi level and bands further away from the Fermi level. Again, this assumption is made mainly to keep the language simple. In principle, a clear energetic separation of the bands does not appear to be necessary in this formalism. 

Let us start with the full theory, including all electronic degrees of freedom.
The original partition function shall be given by the functional integral
\begin{equation}
Z = \int D\psi  \, e^{\frac{1}{2}\psi C^{-1} \psi - V(\psi)}  \label{Z}  \, . 
\end{equation}
Here, in order to keep the notation lean, we have suppressed all quantum numbers $k=(ik_0,\vec{k},\sigma,b,n)$, i.e. Matsubara frequency $ik_0$, wavevector $\vec{k}$, spin projection $\sigma= \pm 1/2$, band index $b$ and Nambu index $n$ for barred or unbarred Grassmann fields $\psi$. Correspondingly, the propagator $C$ is invertible matrix in this super-index space and $\psi C^{-1} \psi$ is a bilinear from involving summations over all quantum numbers. In the case of translation symmetry an spin-rotational invariance, only the part of the propagator that connected barred and unbarred fields with otherwise same quantum numbers will be nonzero, and reads
\begin{equation}
\label{ }
C (k) = \frac{1 }{ik_0 - \epsilon(\vec{k},s,b)  }Ê\, .
\end{equation}
The bare interaction $V(\psi)$ is assumed quartic in the fermion fields and to obey the lattice symmetries, and involves various summations as well.

The underlying strategy for the RG with respect to the band energy is then to divide the free electron propagator $C$ into $C= D_s + C_s$ (see, e.g., Ref.~\onlinecite{salmhoferbook}). $D_s$ is the free propagator for is the low-energy part, which is zero the high energy modes above scale $s$. This can be achieved by writing 
\begin{equation}
\label{ }
D_s (k) = \frac{\Theta \left[s- |\epsilon(\vec{k},s,b)| \right] }{ik_0 - \epsilon(\vec{k},s,b)  }Ê\, .
\end{equation}
Here we write a step function $\Theta \left[s- |\epsilon(\vec{k},s,b)| \right] $ to restrict the propagator sharply on the low-energy modes, but other, smoother choices work similarly.
$C_s$ is the propagator for the high-energy part. 
\begin{equation}
\label{ }
C_s (k) = \frac{\Theta \left[ |\epsilon(\vec{k},s,b)| -s \right]  }{ik_0 - \epsilon(\vec{k},s,b)  }Ê\, .
\end{equation}

\begin{figure}
 \begin{center}
 \includegraphics[width=.75\linewidth]{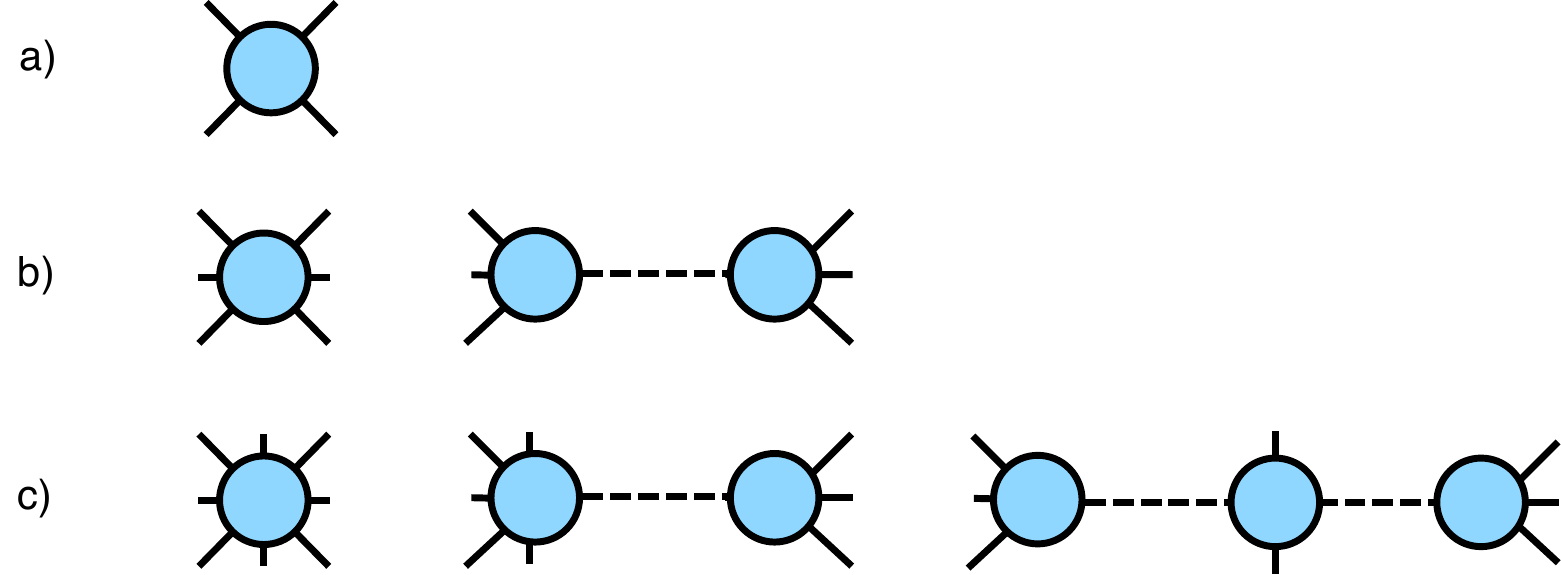}
 \end{center}
 \caption{The interaction terms in the low-energy effective action $S_{\mathrm{eff}}$  obtained from integrating out the high-energy modes above scale $s$, a) is the two-particle interaction ${\cal V}_s^{(4)}= S_{\mathrm{eff}}^{(4)}$, b) the three-particle interaction ${\cal V}_s^{(6)}=S_{\mathrm{eff}}^{(6)}$  and c) the four-particle interaction term ${\cal V}_s^{(8)}= S_{\mathrm{eff}}^{(8)}$. The circles denote the one-particle-irreducible (1PI) vertices $\gamma_{s}^{(n)}$ from the high-energy problem that include all types of 1PI-corrections with internal high-energy lines above scale $s$ only. The tree diagrams in b) on the right and c) in the middle and on the right have single (full) high-energy propagators as connections.}
\label{S468}
\end{figure}

The Gaussian measure splitting formula (see, e.g., Ref.~\onlinecite{salmhoferbook}) leads to (using new Nambu fields $\lambda$ for the low-energy fields and $\eta$ for high-energy fields)
\begin{equation}
Z = \int D\lambda \, e^{\frac{1}{2}\lambda D_s^{-1} \lambda }Ê\int d\eta \, e^{\frac{1}{2}\eta C_s^{-1} \eta} \, e^{-V(\lambda + \eta)}  \label{split} 
\end{equation}
The second integral is (up to normalization with the free partition function $Z_0$ of the high-energy modes) the exponential of Polchinksi's effective action\cite{salmhoferbook,metznerreview,enss,polchinski} ${\cal V}_s (\lambda)$ for the theory at scale $s$, defined by
\begin{equation}
\label{ }
e^{-{\cal V}_s (\lambda)} = Z_0^{-1} \int d\eta \, e^{\frac{1}{2}\eta C_s^{-1} \eta} \, e^{-V(\lambda + \eta)}  
\end{equation}
The expansion coefficients of ${\cal V}_s (\lambda)$ in powers of the low-energy fields $\lambda$ are the connected $C_s$-amputated $m$-point correlation functions, ${\cal V}_s^{(m)}$. Alternatively, these can be obtained by constructing tree diagrams out of the one-particle irreducible (1PI) vertices of the high-energy theory above scale $s$, with full high-energy propagators on connected them. In Fig. \ref{S468} we show the first terms corresponding to two-particle, three-particle and four-particle interactions. Each external leg of these expansion coefficients with quantum numbers $k$ carries a full high-energy propagator $G_s(k)$ divided by a bare high-energy propagator $C_s(k)$. While in principle this dressing may contain interesting physics, we will fully ignore self-energy corrections in the following and will direct our full attention on the interaction terms. In this approximation the external legs are fully amputated, as for the 1PI vertices.

This way, the effective action $S_s(\lambda)$ of the 'low-energy'-fields $\lambda$ can be obtained,
\begin{equation}
Z / Z_0= \int d\lambda \, e^{\frac{1}{2}\lambda D_s^{-1} \lambda }Ê \, e^{-{\cal V}_s (\lambda)}  =  \int d\lambda  \, e^{-S_s(\lambda)} \, .  \label{effaction} 
\end{equation}
This formula is a well-defined starting point for a detailed analysis of the low-energy problem, not only using fRG but also many other many-body approaches.  The effect of the higher-energy degrees of freedom is contained in the expansion coefficients of ${\cal V}_s (\lambda)$, i.e. the amputated connected correlation functions at scale $s$.
The $\lambda$-field propagators in the remaining low-energy problem are bounded from above, i.e. live exclusively in the low-energy sector.  In any perturbative treatment, only low-energy modes contribute on the internal lines of the diagrams. 

The important point to notice  (although maybe obvious to many readers) is that in general, the expansion of the effective (inter-)action  ${\cal V}_s (\lambda)$ does not stop after the fourth power in the fields $\lambda$. In this regard, the effective theory is not necessarily simpler than the original one, and we will soon come back to the information hidden in the terms of order higher than four.

In order to be more clear, imagine a two-band situation, where $C_s$ describes the upper $\eta$-band ('upstairs') above the Fermi level and $D_s$ the lower '$\lambda$-band' closer to the Fermi level ('downstairs'). We are interested in downfolding the model onto the $\lambda$-band, i.e. we can think of scale $s $ as division line between the two bands.
If we use Eq. \ref{effaction}, the renormalizations of the parameters in the $\lambda$-band action will be given by diagrams with only $\eta$-modes in the upper band on the internal lines. The contributions with all particles in the low-energy sector will be collected later in the treatment of the low-energy model. 

This is fine, and may describe some important effects. However, we might worry where the 'mixed' diagrams renormalizing the four-point functions with one internal line in the high-energy sector and one in the low-energy sector have gone. As mentioned before, these diagrams are potentially more important than those with two internal lines upstairs, as they have a smaller energy denominator. in the cRPA formalism, these mixed diagrams are summed up to infinite order, together with the particle-hole bubbles with two intermediate lines upstairs.
In principle, these diagrams are not lost in the RG approach. The three-particle interaction term $S_{\mathrm{eff}}^{(6)}$ at scale $s$ is obtained in parts by a tree diagram made from two 1PI-4pt-vertices $\gamma^{(4)}_s$ with a high-energy propagator as connection (see right term in Fig. \ref{S468} b)). Now, in the perturbative treatment of this low-energy model, two of the external lines of the $S_{\mathrm{eff}}^{(6)}$ will get connected by a low-energy propagator. This means we have a low-energy propagator and a high-energy propagator joining the two four-point vertices (see right term in Fig. \ref{S468cRPA}). This reconstructs the looked-for mixed diagrams. But obviously, we have to include the higher order interaction$S_{\mathrm{eff}}^{(6)}$ to capture these effects. This appears to be a major complication, as most many-body techniques and also the common wisdom on many-fermion systems are focussed on two-particle interactions only, and inclusion of terms beyond this order lead to severe complications. Below we will see that fortunately there is workaround for this problem, and relevant contributions due the six-point term can be recovered without having to compute $S_{\mathrm{eff}}^{(6)}$ explicitly.

\subsection{cRPA as partial resummation}
The sketch of the RG approach to the effective theory made clear that in order to retrieve the mixed particle-hole terms summed in the cRPA, one apparently has to include higher order terms in the effective action. Fortunately, it is not difficult to see that the cRPA basically resums the simplest contributions of such higher order vertices to the two-particle interaction. Hence at least a part of these higher-order interactions is captured. This way one circumvents to treat a low-energy theory with more than two-particle interactions.

\begin{figure}
 \begin{center}
 \includegraphics[width=.75\linewidth]{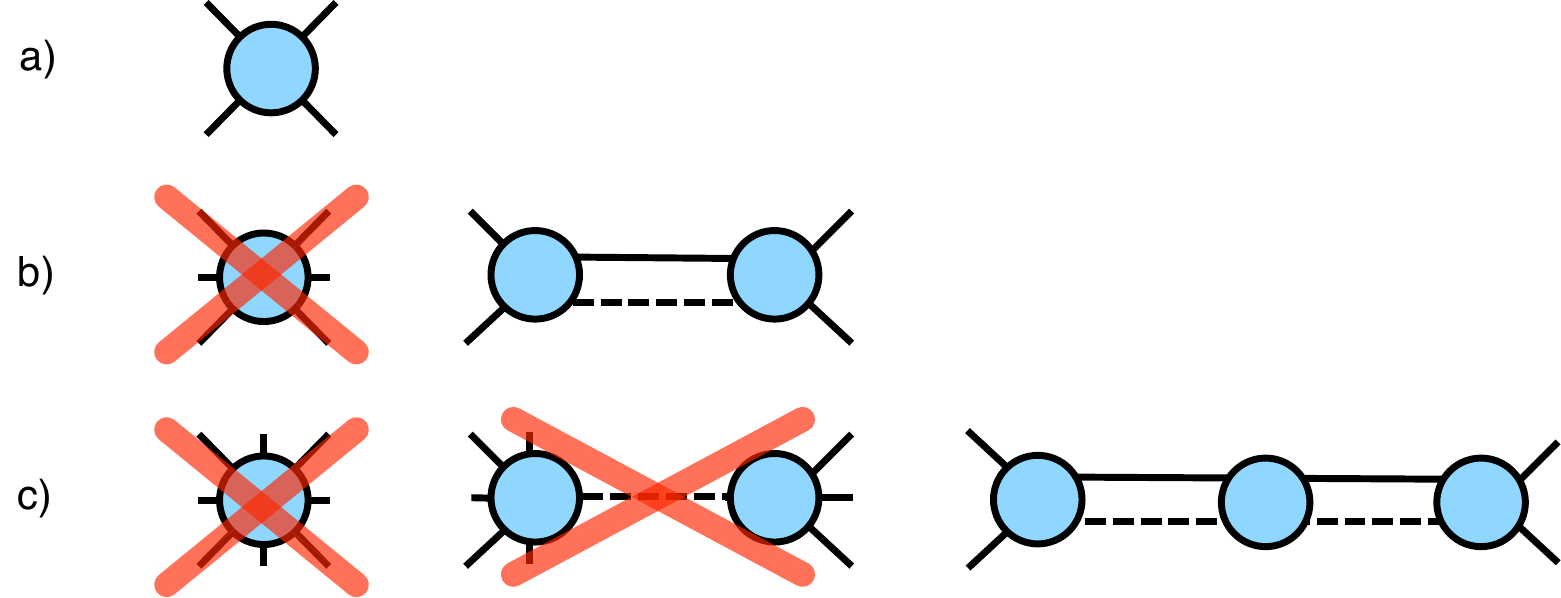}
 \end{center}
 \caption{Resummation of contributions from higher order terms in the effective action that recovers the cRPA. Besides the four-point term in the effective action in a), also specific contractions of the higher-order terms $S_{\mathrm{eff}}^{(m)}$ with four remaining low-energy legs are kept. The crossed out terms are disregarded on this level. The right term of b) reconstructs the second order mixed term in the cRPA, plus other particle-hole corrections and particle-particle diagrams, while the right term of c) brings in the third-order corrections. Including analogous contributions from $S_{\mathrm{eff}}^{(m)}$ with $m>8$ and only keeping the RPA-type particle-hole terms allows one recover the cRPA summation to infinite order.}
\label{S468cRPA}
\end{figure}

In order to understand this construction, let us again consider the first contributions to the interaction terms in the effective action, in Figs. \ref{S468} and then in Fig. \ref{S468cRPA}.  Now, imagine that we only want to keep two-particle interactions with four external legs like the first term a), the 1PI-four-point vertex, as interaction terms in our effective theory. Dropping all terms except a) is the simplest approximation, but this drops all mixed terms and interband particle-hole pairs. If we kept the higher-order terms in the low-energy theory below the separating scale $s$ , and treated them perturbatively, additional  two-particle interactions would be generated by joining two of the external low-energy legs of the interaction terms by low-energy propagators. This either generates tadpoles, if the two legs are at the same 1PI vertex, or one- and more-loop diagrams with two and more propagators, if the two legs are at different vertices. If we only allow to connect legs at neighboring vertices with low-energy lines, we get mixed diagrams like the ones on the right hand of b) and c) in Fig. \ref{S468cRPA}. The right term of Fig. \ref{S468cRPA} b) reconstructs the second order term in the cRPA, while the right term of c) brings in the third-order correction (and much more, because not only the direct particle-hole terms are reconstructed this way).  If we now only keep the appropriate particle-hole terms that build up  the RPA series, and do the same for all higher-order interactions $S_{\mathrm{eff}}^{(m)}$ of order $m > 8$, we sum up the missing mixed particle-hole diagrams from the cRPA to infinite order.

The corresponding bubble sums with two high-energy lines in the individual bubbles are already summed in the 1PI four-point vertex, i.e. the two-particle-interaction 
\begin{equation}
S_{\mathrm{eff}}^{(4)} = \frac{1}{2} \sum_{k,k',q \atop s,s'}  V_{\mathrm{eff}}^{(4)} (q) \bar\lambda_{k+q,s} \bar\lambda_{k'-q,s'} \lambda_{k',s'} \lambda_{k,s} \, , \end{equation} 
 in the effective theory before resumming the higher-order terms. To get the strict correspondence to cRPA, we should also restrict the construction of the effective two-particle interaction $S_{\mathrm{eff}}^{(4)}$ to the RPA-type diagrams only, i.e. write $S_{\mathrm{eff}}^{(4),RPA}$, or 
 \begin{equation}
S_{\mathrm{eff}}^{(4),RPA} = \frac{1}{2} \sum_{k,k',q \atop s,s'}  V_{\mathrm{eff}}^{(4),RPA} (q) \bar\lambda_{k+q,s} \bar\lambda_{k'-q,s'} \lambda_{k',s'} \lambda_{k,s} \, , \end{equation} 
with  \begin{equation}
\label{ }
V_{\mathrm{eff}}^{(4),RPA} (q)  = \frac{v(q) }{1 - P_{\mathrm{high},s} (q) v(q)} 
\end{equation}
using the bare interaction $v(q)$ and the polarization in the high energy bands (now dropping self-energy corrections), 
\begin{equation}
\label{ }
P_{\mathrm{high},s} (q) = 2 \sum_k C_s(k) C_s(k+q) \, .
\end{equation}
$C_s (k)$ is the bare high-energy propagator  that lives on scales above $s$.  In our notation, the polarization sums are negative in the limit of frequency transfer $iq_0 \to 0$. 

Furthermore, if we only allow for this specific particle-hole diagrams that form the RPA, the higher-order 1PI vertices $\gamma_s^{(n>4)}$ will be zero, and the higher-order effective interactions will be just given by specific tree diagrams with bare high-energy propagators, which we call $S_{\mathrm{eff}}^{(n),RPA}$.

In total we get for the effective two-particle interaction, after including these RPA corrections due to the higher order interaction terms, 
\begin{eqnarray}
\nonumber
V_{\mathrm{eff}}^{(4),cRPA}  (q) &=&  V_{\mathrm{eff}}^{(4),RPA}(q)  - V_{\mathrm{eff}}^{(4),RPA} (q) P_{\mathrm{mixed},s} (q) V_{\mathrm{eff}}^{(4),RPA} (q) \\  &&\qquad + \quad V_{\mathrm{eff}}^{(4),RPA} (q) P_{\mathrm{mixed},s} (q)  V_{\mathrm{eff}}^{(4),RPA} (q) P_{\mathrm{mixed},s} (q) V_{\mathrm{eff}}^{(4),RPA} (q) - \dots \, . 
\end{eqnarray}
Here the first term is the direct two-particle interaction without mixed-diagram contributions, the second term comes from contracting the tree for $S_{\mathrm{eff}}^{(6),RPA}$, and the third term is obtained from contracting the neighboring vertices in the two-branch tree for $S_{\mathrm{eff}}^{(8),RPA}$.  The minus signs are chosen to according to the usual diagram rules that also apply in this case with cutoffs. The polarization bubble is given by
\begin{equation}
\label{ }
P_{\mathrm{mixed},s} (q) = 2 \sum_k \left[ D_s(k) C_s(k+q) + C_s(k) D_s(k+q)  \right]
\end{equation}
with the bare low-energy propagator $D_s(k)$ and the high-energy propagator $C_s(k)$ that lives on scales above $s$, in this approximation without self-energy corrections.
 
Obviously, this sum for $V_{\mathrm{eff}}^{(4),cRPA}  (q)$ can be extended to all higher $S_{\mathrm{eff}}^{(n),RPA}$, such that we can sum the geometric series to obtain
\begin{equation}
\label{ }
V_{\mathrm{eff}}^{(4),cRPA}  (q) =  \frac{V_{\mathrm{eff}}^{(4),RPA}(q) }{1 -  P_{\mathrm{mixed},s} (q) V_{\mathrm{eff}}^{(4),RPA}(q) } \, . 
\end{equation}
Furthermore, it is not difficult to see that the two geometric series summing high-energy and mixed polarizations can be combined. This gives in terms of the bare two-particle interaction
 \begin{equation}
\label{ }
V_{\mathrm{eff}}^{(4),cRPA}  (q) =  \frac{V^{(4)}(q) }{1 - \left[ P_{\mathrm{mixed},s} (q) +  P_{\mathrm{high}} (q)  \right] \, V_{\mathrm{eff}}^{(4),RPA}(q) } \, ,
\end{equation}
which is the same as the cRPA result stated in Eq.~\ref{vcrpa}. This completes the identification of the cRPA as a partial resummation of higher-order interaction terms in the effective theory, however with a specific choice that only particle-hole diagrams that can be summed as RPA series are included.
As stated in the introduction, there is no hard reason why other mixed or high-energy diagrams should give much smaller corrections. Hence it would be interesting to try a parquet summation of all one-loop diagrams (i.e. the other particle-hole terms corresponding to vertex corrections and to crossed diagrams, as well as the particle-particle diagrams).
RG schemes like the one for the one-particle irreducible (1PI) vertices are able to do this without the actual need of writing down higher-order terms. They sum all one-loop diagrams to infinite order, but in the usual form they do not resum the higher-order interactions with $m>4$.

\subsection{Relation to Wick-ordered correlation functions}
We will now toward adapting the RG scheme for the four-point  (two-particle) interaction vertices so as to include the missing mixed diagrams and to perform an extended resummation of the higher-order interactions. 
Here, one is aided by the observation that the resummation just described for the cRPA is basically (not exactly, due to the neglected terms) equivalent to the relation between so-called Wick-order correlation functions, generated by a Wick-ordered functional ${\cal W}_s (\lambda)$  and the amputated connected correlation functions, generated by ${\cal V}_s (\lambda)$. This relation reads\cite{wieczerkowski,salmhoferbook,enss}
\begin{equation}
\label{wickorder}
{\cal W}_s (\lambda)  = e^{\Delta_{D_s}} \, {\cal V}_s (\lambda) \, ,
\end{equation}
where (now suppressing bars on the field again, and working with a matrix propagator $D_s$)
\begin{equation}
\label{ }
\Delta_{D_s} = \sum_{k,k'}  \frac{\delta}{\delta \lambda_k} D_s(k,k')  \frac{\delta}{\delta \lambda_{k'}} \,
\end{equation}
is a functional Laplacian which takes away two low-energy fields of an amputated correlation function and then joins the two vacant external legs with a low-energy propagator. The Wick-ordered functional ${\cal W}_s (\lambda)$ obtained in Eq.~\ref{wickorder} can again be expanded in a power series in the fields $\lambda$, with Wick-ordered correlation functions as expansion coefficients. The Wick-ordered functions differ in their physical content from the amputated connected correlation functions. In particular, they include to looked-for mixed diagrams. 

\begin{figure}
 \begin{center}
 \includegraphics[width=.75\linewidth]{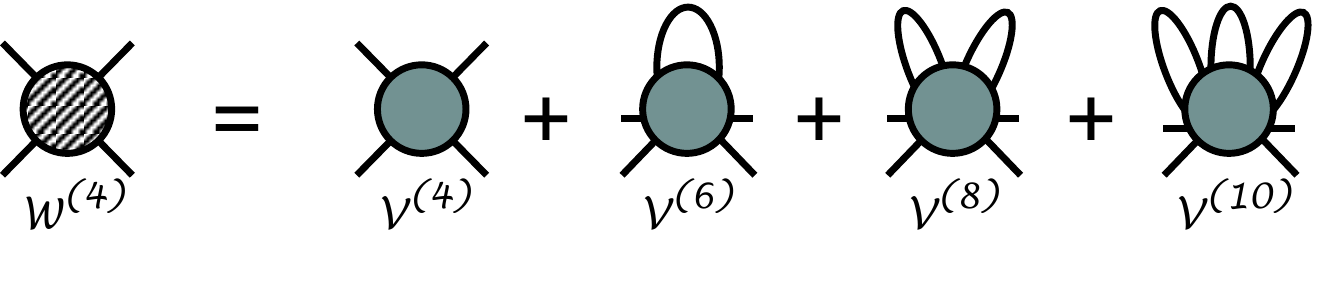}
 \end{center}
 \caption{Relation between the fourth-order term ${\cal W}_s^{(4)}$  in the expansion of the Wick-ordered generating functional defined in Eq.~\ref{wickorder} and the amputated connected correlation functions ${\cal V}_s^{(m)}$. The closed lines are bare low-energy propagators.}
\label{wickpol}
\end{figure}

This can be seen most easily if we focus on the fourth-order Wick correlation function, ${\cal W}_s^{(4)}$. The exponential of the Laplacian generates an infinite series of terms of fourth order in $\lambda$, starting with the normal fourth-order term ${\cal V}_s^{(4)}$ (which is the same as the 1PI-four point vertex $\gamma_s^{(4)}$ when we neglect self-energies) but then contains as next term a singly $D_s$-contracted ${\cal V}_s^{(6)}$, a doubly contracted ${\cal V}_s^{(8)}$ and so on. 
The diagrammatic expression for the relation between the Wick-ordered four-point correlation function and the amputated correlation functions is shown in Fig. \ref{wickpol}. Note that in a theory with bare two-particle-interactions only, building up a $2m$-point amputated correlation function from the bare interaction then requires at least $m-2$ high-energy propagators for the necessary tree diagrams (the 1PI-parts expressed in the bare interaction have even more internal high-energy lines). In forming the four-point Wick function, $2(m-2)$ legs are contracted with $m-2$ low-energy propagators. Hence, in terms of the bare interaction, the perturbation series for the four-point Wick function contains diagrams that have at least as many internal high-energy lines as low-energy lines. In this sense the Wick interaction is somehow half-way between only capturing high-energy corrections and the full perturbation series on all scales. 
 
The cRPA can be retrieved in this construction by again neglecting all 1PI vertices higher than $\gamma_s^{(4)}$ and all contractions that do not correspond to those diagrams that occur in the RPA-type mixed particle-hole bubble chain. But the full Wick four-point function contains many other corrections. In the diagram class with as many high-energy as low-energy lines these are particle-hole vertex corrections and particle-particle diagrams with one high-energy and one low-energy line. Regarding diagrams with only high-energy lines, only the RPA particle-hole diagrams going into ${\cal V}_s^{(4)}$ are kept by the cRPA, but the full Wick function also contains the corrections of the other one-loop terms to ${\cal V}_s^{(4)}$, and those higher-energy corrections that build up the 1PI parts of ${\cal V}_s^{(>4)}$.

Thus we arrive at the following proposal. If one does not want to retain the higher-order tail of the effective low-energy action explicitly, but still wants to keep an important part of their feedback such as the cRPA terms and possibly more, one should use the fourth-order Wick correlation function as effective interaction term. Then one can truncate after the fourth order in the low-energy fields, and the theory remains tractable.

\subsection{RG scheme for the Wick-ordered correlation functions}
We have just argued that instead of the amputated connected correlation function of fourth order, the Wick-ordered four-point function should be used in the truncated effective action. This prompts the question how the Wick-ordered correlation functions can be computed efficiently. Here one can take advantage of the corresponding RG flow equations that were developed by Wieczerkowski\cite{wieczerkowski} for scalar field theories and by Salmhofer\cite{salmhofer_wick} for fermions. In the context of correlated fermions, these Wick-ordered RG equations were applied in the proof of Fermi liquid behavior in two dimensions for sufficiently small interactions and benign dispersions\cite{salmhofer_wick}, and for instability-\cite{halboth} and self-energy-\cite{rohe} studies in the two-dimensional Hubbard model on the square lattice. In the same context, the Wick-ordered scheme has already been used as a tool to determine the effective interactions for a mean-field study\cite{reiss}.

In the case of spin-rotational invariance, any antisymmetric four-point function can be expressed in terms of a coupling function $V_s(p_1,p_2,p_3,p_4)$ where the spin projection $\sigma$ of the first incoming particle with other quantum numbers $p_1$ euqals that of the first outgoing particle $p_3$, and spin projection $\sigma'$ is the same for the second incoming particle $p_2$ and the second outgoing one $p_4$.
On the level where self-energy corrections are neglected, and no higher-order Wick function beyond the fourth-order term are considered, the flow equations for the fourth-order Wick interaction function $V_s(p_1,p_2,p_3,p_4)$ read \cite{halboth}
\begin{align}
\label{dgl_v}
&\frac{d}{ds} V_s ( p_1,p_2,p_3,n_4 ) =\tau_s^{PP}+\tau_s^{PH,d}+ \tau_s^{PH,cr},
\end{align}
with the particle-particle channel
\begin{align}
\label{pp}
\tau^{PP}_s ( p_1,p_2,p_3,n_4 )= - T\int dp \sum_{n'} V_s ( p_1,p_2,p,n' ) 
L_s ( p,q_{PP} ) V_s ( p,q_{PP},p_3,n_4 ) \, , 
\end{align}
and the direct particle-hole channel
\begin{align}
\tau&_s^{PH,d}( p_1,p_2,p_3,n_4) = -T \int dp \sum_{n'} \Bigl[ 
 \underline{-2V_s(p_1,p,p_3,n')  L_s(p,q_{PH,d} ) V_s(q_{PH,d},p_2,p,n_4)}  \notag\\
&+ V_s(p,p_1,p_3,n')L_s(p,q_{PH,d} ) V_s(q_{PH,d},p,n_4) + V_s(p_1,p,p_3,n') L_s(p,q_{PH,d} ) V_s(p_2,q_{PH,d},p,n_4)    \Bigr] \, . \label{phd} 
\end{align}
Here, the underlined term in the first line on the right hand side can be identified as RPA correction, it also carries the factor of two from the internal spin sum. Only keeping this term and dropping all the other terms on the right hand side of Eq.~\ref{dgl_v} reproduces the RPA diagram series (the precise mode content of the diagrams depends on the cutoff used in the RG, here it corresponds to cRPA). The other two terms in Eq.~\ref{phd} are vertex corrections.
Finally there is the crossed particle-hole channel
\begin{align}
\label{phcr}
\tau_s^{PH,cr}&( p_1,p_2,p_3,n_4 )= - T \int dp \sum_{n'} V_s(p,p_2,p_3,n')  L_s(p,q_{PH,cr} ) V_s(p_1,q_{PH,cr},p,n_4), 
\end{align}
where $q_{PP}=(-\vec{k}+\vec{k}_1+\vec{k}_2;-w+w_1+w_2;n')$, $q_{Ph,d}=(\vec{k}+\vec{k}_1-\vec{k}_3, \omega +\omega_1-\omega_3;n')$, $q_{Ph,cr}=(\vec{k}+\vec{k}_2-\vec{k}_3,\omega+\omega_2-\omega_3,n')$ are the quantum numbers of the second loop line, $p=(\vec{k},\omega,n)$ are those of the first line. 
The loops are given by
\begin{equation}
L^s(p,p')= \frac{d}{ds}Ê\left[ D_{s} (p) D_{s} (p') \right] \, ,
\end{equation}
with the bare low-energy Green's function defined to be nonzero only {\em below}Ê the scale $s$ (at least if a sharp cutoff is used), 
\begin{equation}
\label{ }
D_{s} (p) = \frac{\Theta \left[ s- |\epsilon(\vec{p})| \right] }{ ip_0 - \epsilon(\vec{p}) } \, . 
\end{equation}
The initial condition at scale $s=W$, the bandwidth of the multiband-system including the higher energy bands, is given by the bare interaction, $V_{s_0} (p_1,p_2,p_3,p_4) = V_{n_1,n_2,n_3,n_4} (\vec{p}_1,\vec{p}_2,\vec{p}_3,\vec{p}_4) $, which can be found, e.g., by computing Coulomb integrals in an already downfolded multiband Wannier representation, or by staying in the full Bloch representation with all bands. 

Integrating the coupled RG equations in Eq. \ref{dgl_v} down to a scale $s$ generates the full parquet sum, i.e.~all diagrams of arbitrarily high order in the bare interactions that can  be obtained from repeated replacing of a bare interaction by one-loop diagrams, either of particle-particle type or the various particle-hole types. When we only integrate from the initial scale $s_0$ down to $s>0$, at least half of the internal lines in these one-loop diagrams have support only on the high-energy sector $s \le |\epsilon_n(\vec{p})| \le s_0$, just as in the direct perturbation expansion of the four-point Wick function discussed above. In particular the pure low-energy corrections that have all internal lines below scale $s$ remain untouched. 

Quite generally, the initial interaction will not be frequency-dependent, and only depend on three wavevectors and four band indices (spin rotational invariance is assumed). Then, in the flow to lower cutoff, the coupling function will acquire an additional dependence on three frequencies. We will use the two incoming frequencies $\omega_1$ and $\omega_2$ and one outgoing frequency $\omega_3$ to capture this dependence, where again the notation is such that the first incoming line 1 and the first outgoing line 3 have the same spin projection. Another choice would be to use the total incoming frequency $\omega_1+ \omega_2$ and two frequency transfers $\omega_1- \omega_3$ and $\omega_2- \omega_3$. 

The cRPA can be recovered by restricting the right hand side of the flow equation to the underlined RPA term, and ignoring all the other contributions. This just sums RPA contributions with one line in the high-energy sector between $s_0$ and $s$, and the other line in the full energy window between $s_0$ and $0$. Then the coupling function will only depend on one frequency transfer $\omega_1- \omega_3$.
This way we have at hand a comprehensive formalism that includes the cRPA as approximation. This gives us an appropriate  framework to compare the changes with respect to cRPA when we include all other one-loop diagrams.

\section{Application and comparison in a simple two-band model} \label{sec:res}
In this section we compare the effective interactions obtained with the Wick-ordered fRG scheme and the cRPA in a simple two-band model. This will mainly give us information on the difference in the degree of the suppression of the repulsion by the screening, and on the differences in the frequency structure in the effective interactions.

\subsection{Model and implementation of fRG scheme}
\begin{figure}
 \begin{center}
 \includegraphics[width=.5\linewidth]{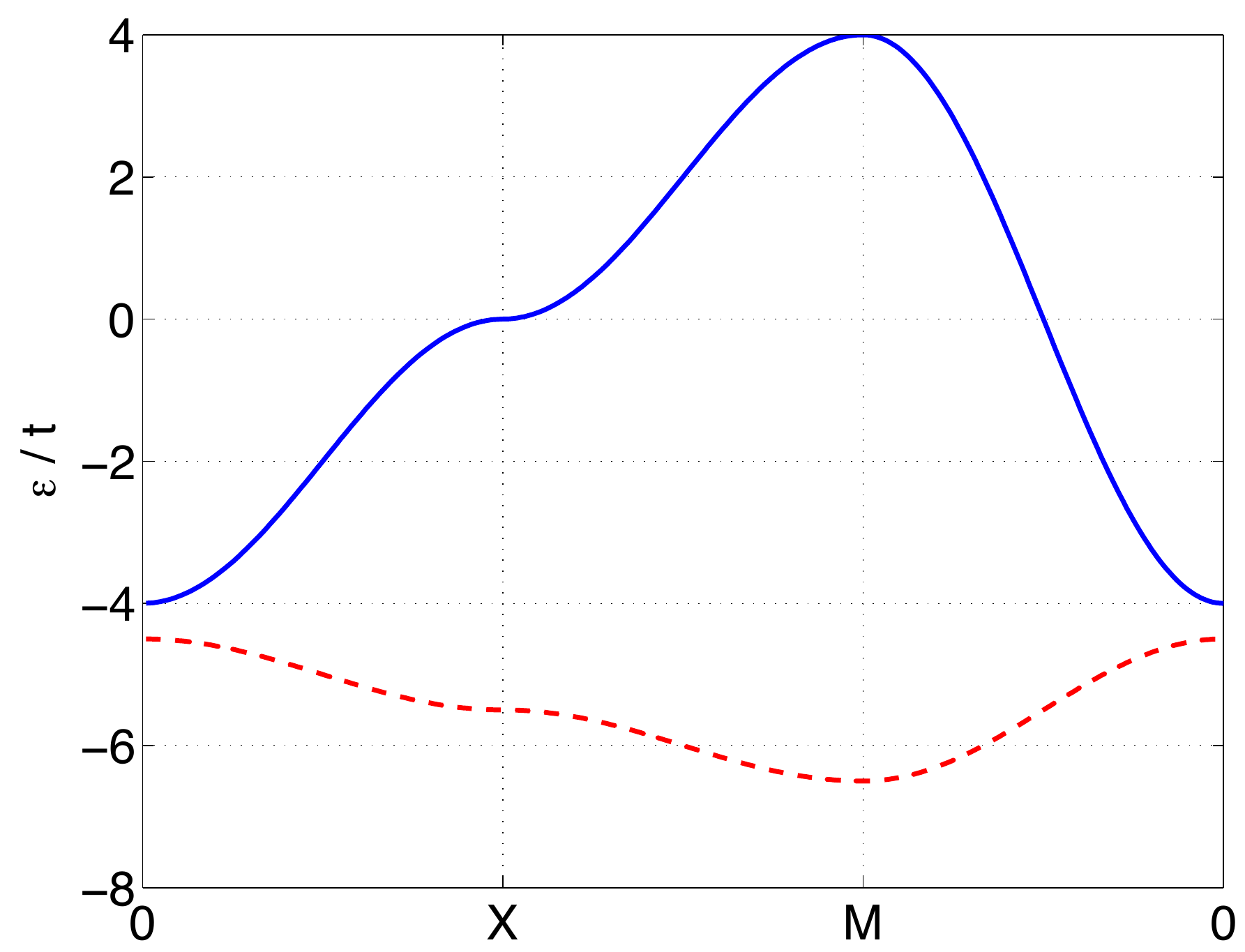}
 \end{center}
 \caption{Dispersion of the two-band model as used for the numerical study in Sec. \ref{sec:res} for $\Delta E =5.5t$, with the other parameters mentioned in the text. The points on the trajectory through the two-dimensional square lattice Brillouin zone are the $0=(0,0)$, $X=(\pi,0)$, and $M=(\pi, \pi)$. The upper band (solid line) is the conduction band for which the effective interactions are computed by integrating out the lower 'valence' band (dashed line), either by cRPA or by Wick-ordered fRG.}
\label{displot}
\end{figure}

The model is given by the Hamiltonian
\begin{equation}
\label{hamilton}
H = \sum_{\vec{k},s,b} \epsilon_b(\vec{k}) c_{\vec{k},s,b}^\dagger c_{\vec{k},s,b} + \frac{U}{2} \sum_{\vec{k},\vec{k}',\vec{q} \atop b_1 \dots b_4, s,s'} c_{\vec{k}+\vec{q},s,b_3}^\dagger c_{\vec{k}'-\vec{q},s',b_4}^\dagger  c_{\vec{k}',s',b_2} c_{\vec{k},s,b_1} \, , 
\end{equation}
with the two bands
\begin{equation}
\label{cband}
\epsilon_c (\vec{k} ) = - 2t (\cos k_x + \cos k_y )  - \mu
\end{equation}
for the two-dimensional conduction band (band index $b=c$), and 
\begin{equation}
\label{vband}
\epsilon_v (\vec{k} ) = - 2t_v (\cos k_x + \cos k_y )  - \Delta E - \mu
\end{equation}
 for the 'valence' band (band index $b=v$) which is in our case $\sim \Delta E$ below the Fermi level. We use the parameters $t_v =-t/4$ and $\Delta E \sim 5t$. The dispersion is shown in Fig. \ref{displot}.
 
For the onsite interaction part we simply use a constant $U=t$ irrespective of the band index. This approximation ignores a lot of quantitative structure, sometimes called {\em orbital make-up}\cite{maier}, that is brought in when the  bare interaction in orbital space is rewritten in band language. We prefer to drop this aspect in the discussion here, as it might complicate the comparison between the two approximation schemes. On technical level, it is straightforward to include more realistic parametrization of the interactions. This also holds with some restrictions for the wavevector or spatial dependence of the interactions, a point which we will comment on in the next section.
 
\begin{figure}
 \begin{center}
 \includegraphics[width=.9\linewidth]{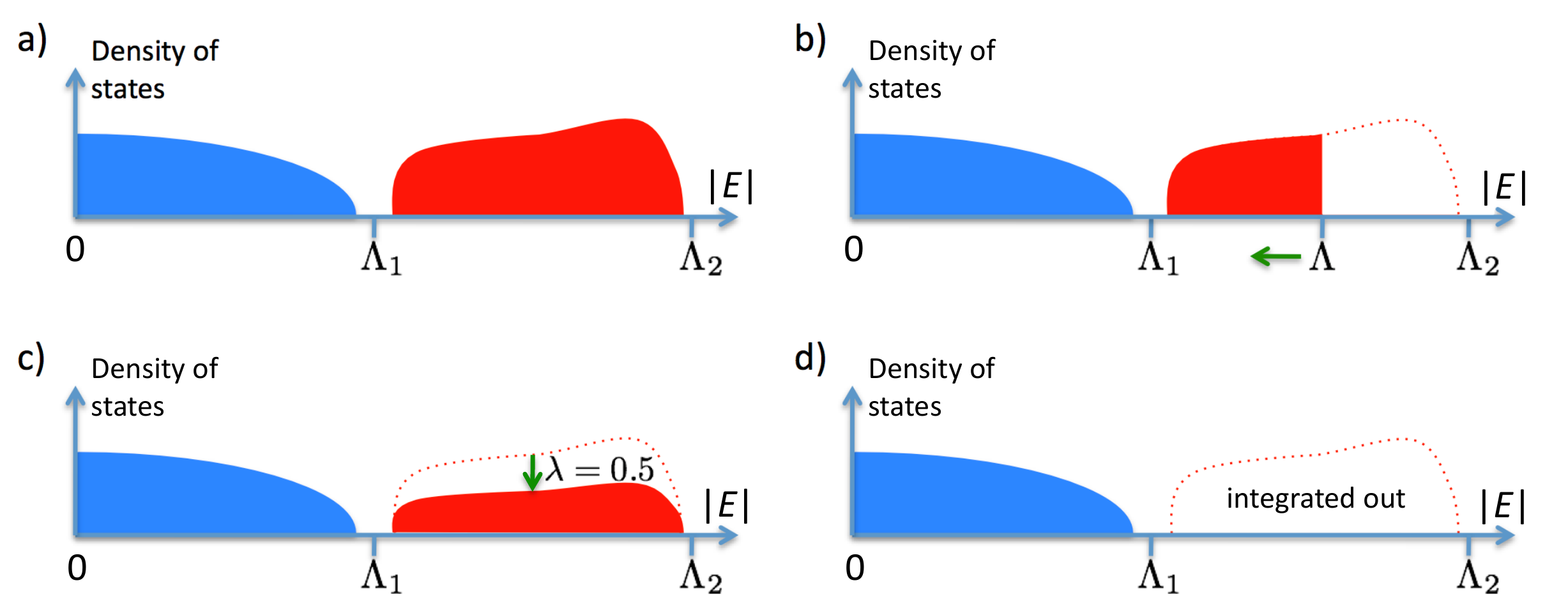}
 \end{center}
 \caption{Illustration of two schemes to integrate out the high energy spectra weight in an RG flow. Figure a) shows the initial condition of both flows (either $\Lambda\ge \Lambda_2$ or $\lambda=0$), with the low-energy degrees of freedom below energy scale $\Lambda_1$ as well as the high-energy degrees of freedom between  $\Lambda_1$ and $\Lambda_2$ active. In b) we show the effect of an energy cutoff. The spectral weight above scale $\Lambda$ has been integrated out, and has effectively been absorbed into the action below scale $\Lambda$. If $\Lambda $ is reduced down to $\Lambda_1$, one obtains the effective one-band model, as shown in d). In c) an intermediate stage of the flat-cutoff scheme used for the numerical comparison is shown. Here, the spectral weight for each mode in the high-energy window between  $\Lambda_1$ and $\Lambda_2$ has become reduced by a factor $1-\lambda$, where $\lambda$ flows from 0 to 1, and the missing fraction $\lambda$ has been absorbed into the low-energy theory. At $\lambda =1$, again the high-energy spectral has been fully absorbed into the low-energy effective theory. This final state, corresponding either to $\Lambda=\Lambda_1$ or to $\lambda=1$, with only the degrees of freedom below $\Lambda_1$ left in the theory, is shown in d). }
\label{cutoffs}
\end{figure}

For the Wick-ordered fRG scheme described in the last section, the standard procedure would be to introduce an energy cutoff that excludes the high-energy modes, which is then lowered from the full bandwidth of the two-band system, say $\Lambda_2$ down to the energy scale where the valence band has been integrated out but the conduction band has not been included, say $\Lambda_1$. This way, the excitations in the valence band are included step by step. Another path to arrive at the same goal is to multiply the propagator in the valence band with a factor $1-\lambda$, and increase $\lambda$ from 0 to 1.  The conduction band propagator remains untouched, i.e. comes with a factor of one in the the loop diagrams where the undifferentiated line is in the low-energy sector (while the differentiated line is alway in the high-energy window). This way the flow from $\lambda=0$ to $\lambda=1$ interpolates between the full model with both bands active to the situation where the propagator in the upper band is switched off, and all diagrams with only conduction band propagators on the internal lines remain unincluded. These are the same  initial and final conditions as for the Wick-ordered flow with energy cutoff flowing from $\Lambda_2$ to $\Lambda_1$. The two cutoff schemes are visualized in Fig. \ref{cutoffs}
This cutoff choice corresponds to a 'flat' cutoff in the band away from the Fermi surface. In single band problems, this flat cutoff flow corresponds to a flow in the bare interaction strength and has been tried out in Refs. \onlinecite{ape}. There it was shown to give equivalent results regarding the interaction-driven instabilities as the conventional momentum-shell schemes.  The flat cutoff flow has one big practical advantage, at least in the simple approximation where self-energy corrections are excluded. Here, one-loop diagrams can be computed once, and are then just rescaled by the factor $(1-\lambda)^2$ (or the $\lambda$-derivative thereof) at a given $\lambda $ in the flow. With the energy-shell cutoff the one-loop diagrams have to be recomputed at every RG step. In general, the optimal cutoff choice, i.e. flat or sharp in energy or combinations, may depend on what one is interested in and on how much numerical effort one is willing to take. We do not expect drastic qualitative differences because the diagram series summed in this RG is non-singular, i.e. the summation should not depend on how the contributing terms are sorted.  

The effective interaction can then be parameterized by a coupling function $V_\lambda ({\vec{k}_i}, {n_i}, {b_i})$, with two incoming wavevectors $\vec{k}_1$ and $\vec{k}_2$, and two outgoing wavevectors $\vec{k}_3$ and $\vec{k}_4$, obeying wavevector-conservation on the lattice.  The real numbers $n_i$ index the fermionic Matsubara frequencies $\omega_{n_i} = (2n_i -1)\pi T$ of these 4 particles, and the $b_i$ are the corresponding band indices.
Spin rotational invariance is assumed, and the convention is such that the spin projection $s$ is the same for the first incoming particle 1 and the first outgoing particle 3, and $s'$ is the spin projection of particles 2 and 4. 

Treating the full wavevector and frequency-dependence of the coupling function to reasonable precision is a difficult task, at least without further approximations or transformations, or without performing heavy numerics. In the bare interactions, there will typically be no retardation, i.e. the initial coupling function is usually frequency-independent, but during the flow, a frequency-dependence will get generated. On the other hand, if one is interested in realistic situations, the wavevector-dependence will usually reflect the long-range part of the Coulomb- interaction, i.e. behave for small $q$ like $1/q^2$ where $\vec{q}$ is the wavevector transfer $\vec{k}_1-\vec{k}_3$. Note however that a major application of the here discussed cRPA is to provide local or almost local Hubbard-type interaction parameters for effective low-energy lattice models with short-ranged interactions. These parameters are wavevector-integrated quantities, and strong wavevector-dependences will be averaged out in the integrals. Hence it should be reasonable to study the case for generic, rather than special or small wavevector transfers, as they should contribute the bulk part to these local parameters. If on the contrary the small-$q$-part would dominate the Hubbard interaction parameters, it would presumably a bad approximation to drop the long-range part of effective interactions.

With this in mind we separate the problem in two parts. First, in this section, we ignore the wavevector-dependence of the initial interaction by simply writing the constant $U$ in the interaction in Eq. \ref{hamilton}. We can then compute the effective interaction in the conduction band and compare between the full RG and cRPA. We will see that the average coupling strength and also the frequency structure come out rather differently. Then, in the next section we discuss, mostly neglecting the frequency structure, what changes can be expected when the initial interaction is of long-range Coulomb-part, and when vertex corrections are less important compared to the RPA contributions.
 
To be more precise we simply discretize the two-dimensional Brillouin zone square in four plaquettes centered at $(\pi/2,\pi/2)$, $(\pi/2, -\pi/2 )$, $ (-\pi/2,\pi/2)$ and $(-\pi/2,-\pi/2)$. In this way we cannot describe anything else that local and nearest neighbor intra- and inter-orbital interactions. We have also made some test runs for a 3$\times$3 discretization, with no qualitative differences regarding the conclusion further below.
 The frequency information is included by computing the coupling function  $V_\lambda ({\vec{k}_i}, {n_i}, {b_i})$, for Matsubara indices $n_i = -N_\omega/2+1, \dots 0 , \dots N_\omega/2$. For higher frequencies on the legs of the loop internal summations, the nearest frequency is used for evaluating the coupling function connecting to this leg. 
Below we present data for $N_\omega=20$, $T=0.1t$ and $T=0.25t$. For $T=0.1t$, the maximal Matsubara frequencies are roughly $\pm 6t$, somewhat larger than the energy gap between the bands. 

\subsection{Numerical results for wavevector-independent interactions}
\begin{figure}
 \begin{center}
 \includegraphics[width=.7\linewidth]{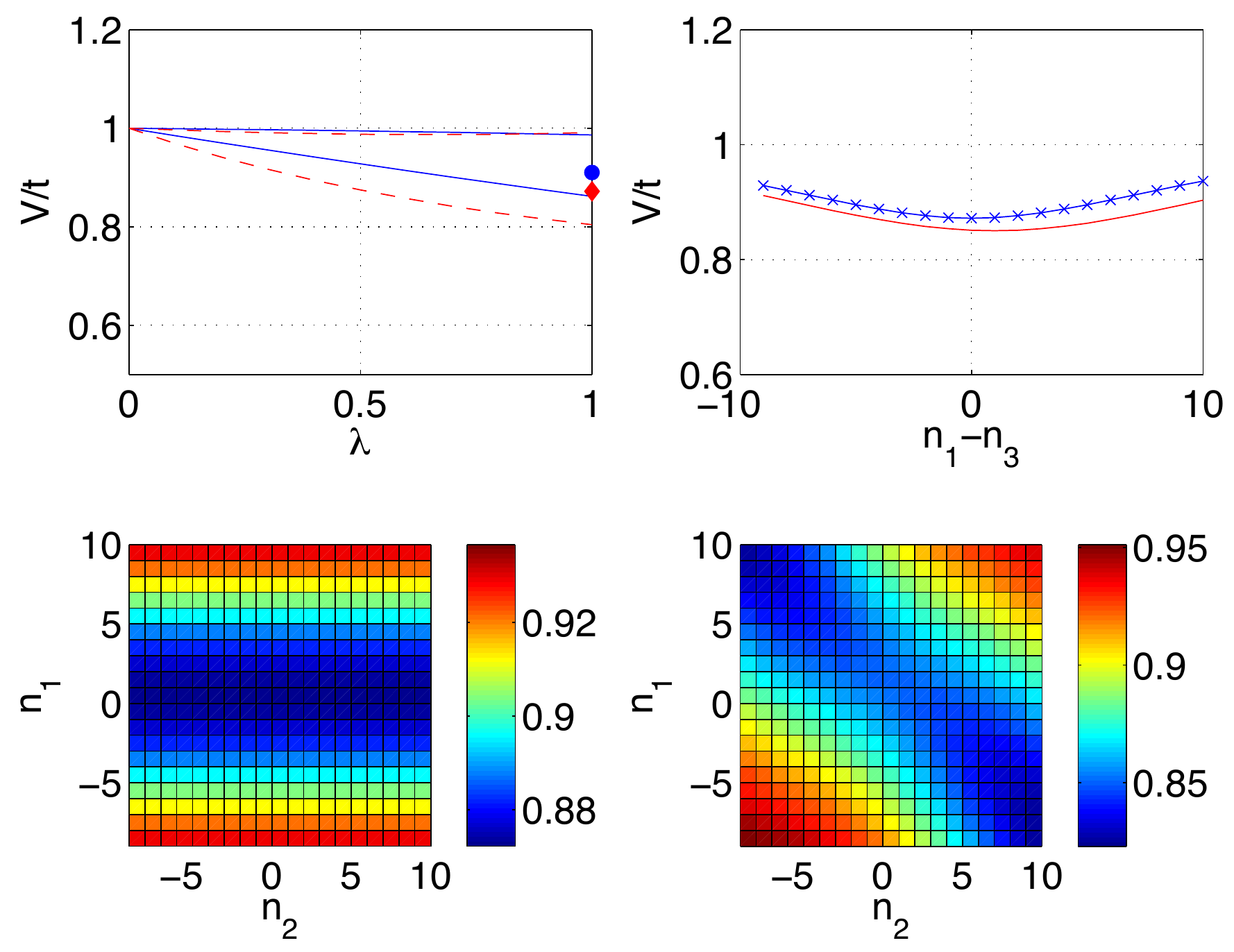}
 \end{center}
 \caption{Comparison of the cRPA and the full fRG for the simple two-band model, obtained for an interaction function $V_\lambda (p_1,p_2,p_3)$ on a $2\times 2$ Brillouin-zone discretization grid  (points $(\pm \pi/2,\pm \pi/2)$) for the wavevector dependence and the 20 lowest Matsubara frequencies for the frequency dependence. The temperature is $T=0.1$ and the chemical potential is $\mu=t$, i.e. the conduction band is more than half filled. The energy separation between the bands is $\Delta E=5.0t$, and the bare interaction is unity, $U=t$, for all frequency-, wavevector- and band-combinations. Upper left plot: Flow of the maximal and minimal components of the effective interaction in the conduction band. The full lines are for cRPA, the dashed lines for the full one-loop fRG. $\lambda=0$ is the initial bare interaction. For increasing $\lambda >0 $, the high-energy window is integrated out successively, for $\lambda=1$ all high-energy modes are absorbed into the low-energy interaction. Upper right plot: Frequency dependence of the effective interaction $V_{\lambda=1} (\omega_1,\omega_2,\omega_3)$ for fixed $\omega_2=-\pi T$ and wavevector transfer $(\pi,0)$. The lower curve is for cRPA, and the upper one for the full fRG. Lower left: Dependence of  $V_{\lambda=1} (\omega_1,\omega_2,\omega_3)$ with $\omega_3$ fixed at $-\pi T$, again for wavevector transfer $(\pi,0)$. The vertical axis contains the first incoming frequency $\omega_1$, and the horizontal axis the second incoming frequency $\omega_2$. The colorbar denotes the change with respect to the bare interaction with magnitude 1. Lower right plot: The same for the full fRG described in the text. }
\label{OFU10_100}
\end{figure}

\begin{figure}
 \includegraphics[width=.7\linewidth]{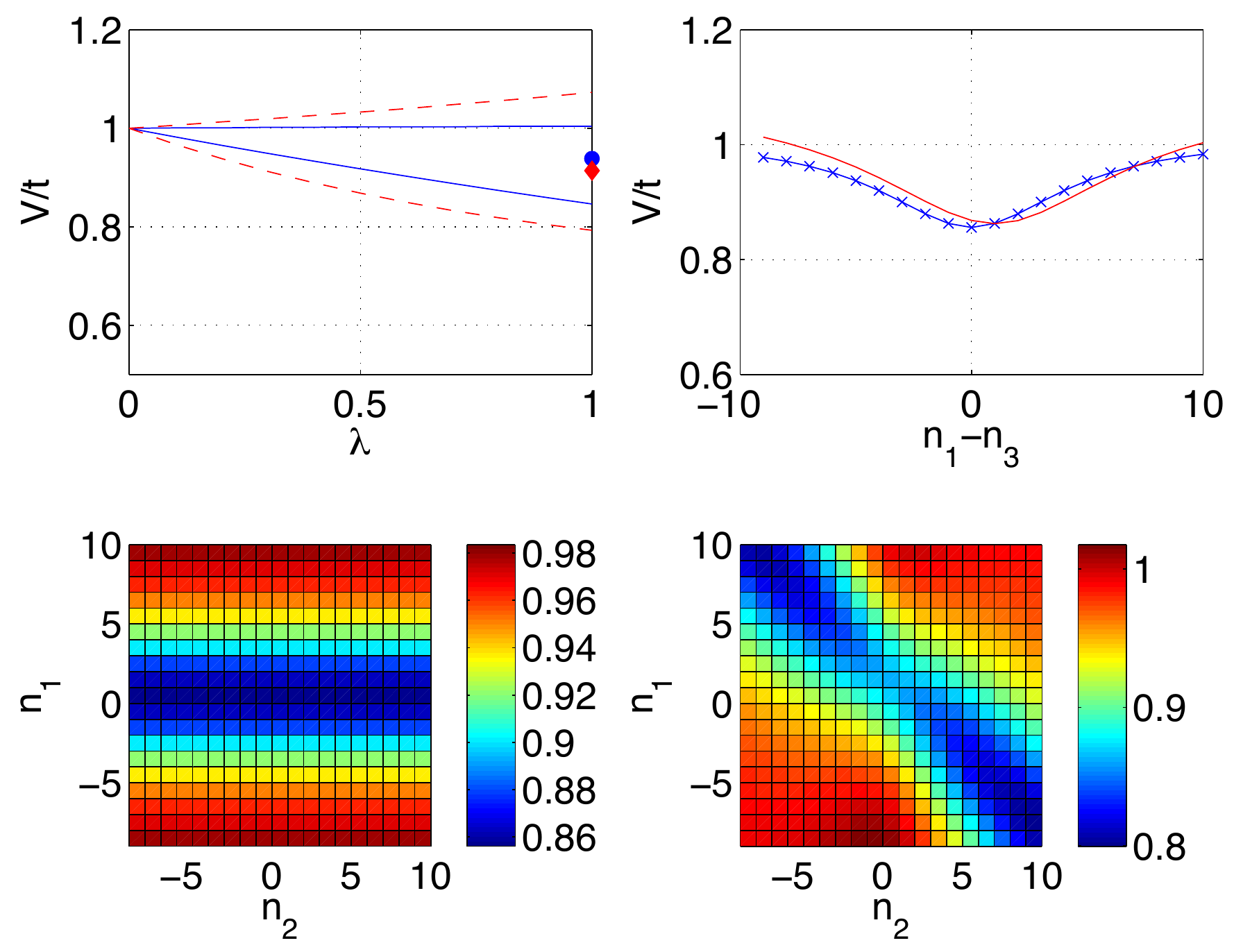}
 \caption{The same as in Fig. \ref{OFU10_100}, but at higher $T=0.25t$. The 10th Matsubara frequency is now $\omega_10 = 19 \pi T \approx 15t$.}
\label{OFU25_100}
\end{figure}

\begin{figure}
 \begin{center}
 \includegraphics[width=.7\linewidth]{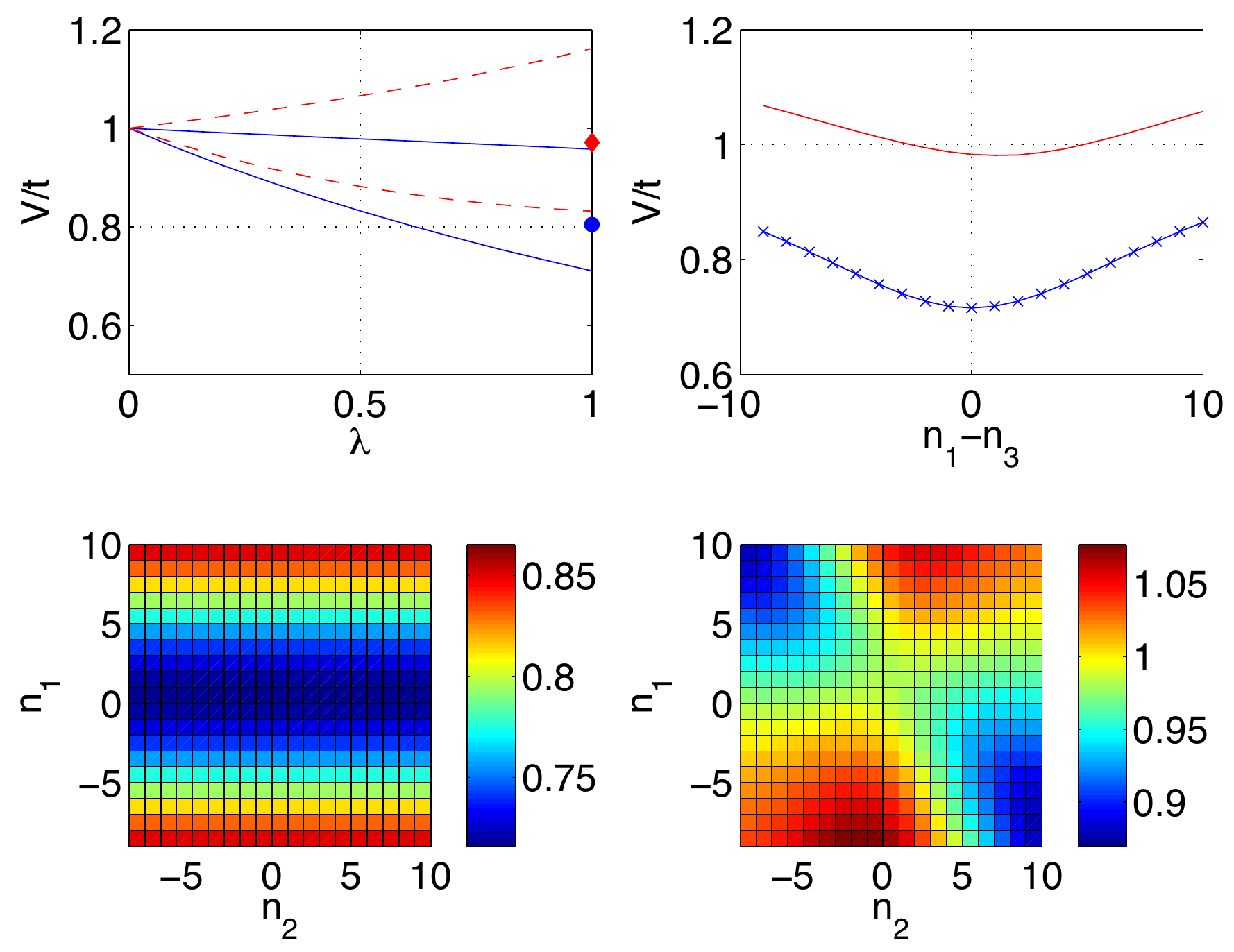}
 \end{center}
 \caption{The same as in Fig. \ref{OFU10_100}, but at less than half-filling, $\mu=-0.5t$, for the conduction band. Now the effective interactions in the full fRG are much less suppressed than in cRPA.}
\label{OFU10_-50}
\end{figure}

Here we describe the results of the numerical study on the simple  two-band model just described.
In Figs. \ref{OFU10_100}, \ref{OFU25_100}, \ref{OFU10_-50}, we show the main characteristics of the data obtained with the Wick-ordered fRG scheme with the full set of one-loop diagrams included and in the approximation where the scheme is reduced to cRPA by dropping the particle-particle-channel, the vertex corrections and the crossed particle-hole channel. 
In the upper left parts of these figures, the flow of the maximal and minimal components of the effective interactions and the average value at the end of the flow are plotted. $\lambda =0$ corresponds to the bare interactions, and $0 < \lambda <1 $ indicates the 'percentage' to which the high-energy band has been absorbed into the effective intra-conduction band interaction. The pair of dashed lines shows the maxima and minimal coupling in the full fRG, while the solid lines are for the cRPA. The final frequency- and wavevector-averaged value of the cRPA is indicated by a blue circle, and by a red diamond for the full fRG. The average reduction is $\sim 10\%$ in both cases. In ab-initio applications of the cRPA, the reduction is usually  stronger, most likely as the bare interaction is stronger on average. Another obvious tuning parameter is the band separation $\Delta E$. Lowering of $\Delta E$ gives stronger renormalizations.
In Figs. \ref{OFU10_100}, \ref{OFU25_100}, which correspond to a conduction band that is more than half-filled, the average reduction of the effective full-fRG interactions is quite be similar than in cRPA. Note however, that the frequency structure of the effective interactions in the full fRG and in cRPA, shown in the lower plots, is quite different. While in the cRPA, the effective interactions only depend on the transfer $\omega_1-\omega_3$, the fRG interactions depend on all three frequencies, and mostly on the total incoming frequency $\omega_1+\omega_2$. Below we analyze this dependence further.

In Fig. \ref{OFU10_-50}, we show analogous results for less than half-filling of the conduction band. This filling opens, in combination with the single filled valence band below the Fermi level, a larger particle-hole phase space. As can be seen in the upper left plot of Fig.  \ref{OFU10_-50}, now the suppression of the effective interactions in the full fRG is somewhat weaker than in cRPA. In the fRG, the reduction at $\lambda=1$ is only a few percent $\sim 3\%$, while the maximal component is actually larger than 1.1, i.e. increased with  respect to the bare value. On the contrary, in cRPA, we find an average reduction down to 80$\%$ of the bare value at $\lambda=1$.  

\begin{figure}

\includegraphics[width=.75\linewidth]{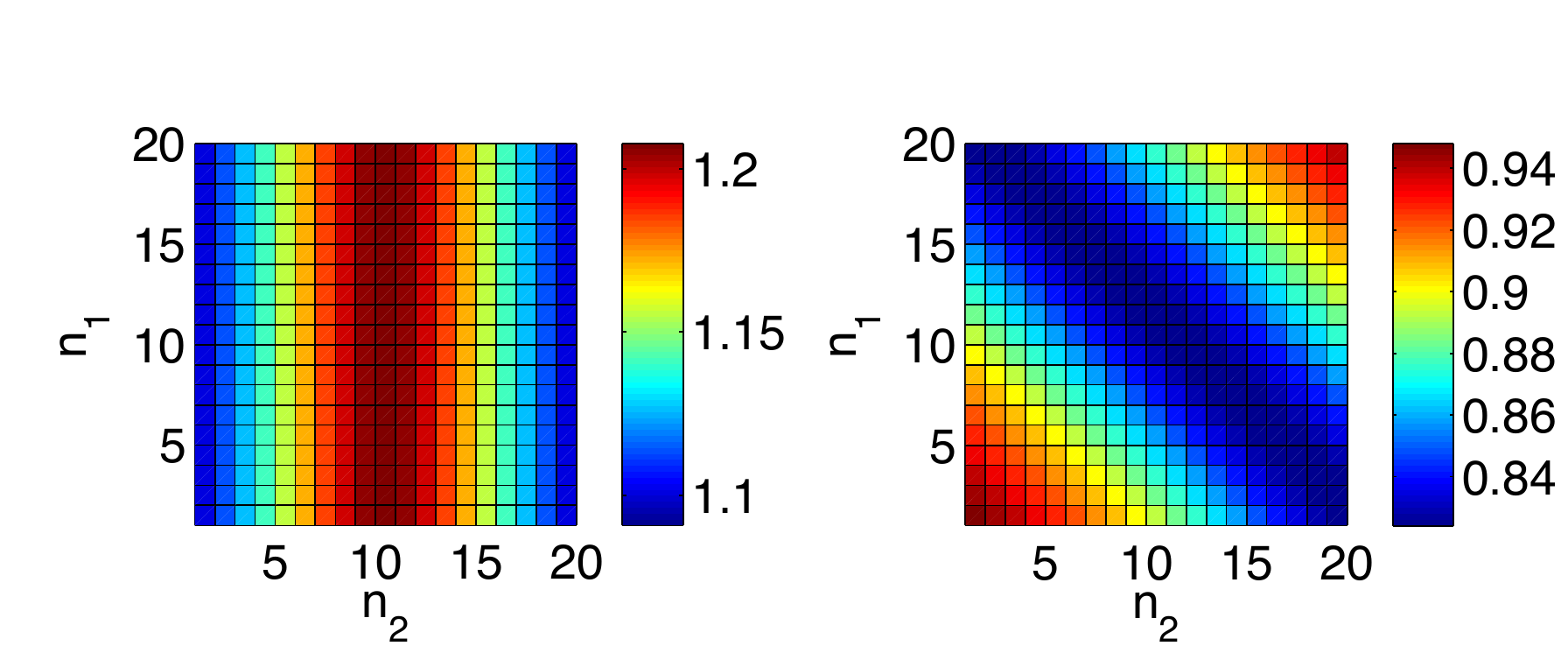}
\caption{Data for the effective interactions $V_{\lambda=1} (\omega_1,\omega_2,\omega_3)$ at $T=0.1t$ and for wavevector transfer $(\pi,0)$, with the right hand side of the fRG flow equation reduced to either the crossed particle-hole channel (left plot) or the particle-particle channel (right plot). The vertical axis contains the first incoming frequency $\omega_1$, and the horizontal axis the second incoming frequency $\omega_2$. $\omega_3$ is fixed at $-\pi T$. The colorbar illustrates the reduction or enhancement with respect to the initial value $V_{\lambda=0} (\omega_1,\omega_2,\omega_3)=t$.}
\label{OFUsc}
\end{figure}

\begin{figure}
 \includegraphics[width=.75\linewidth]{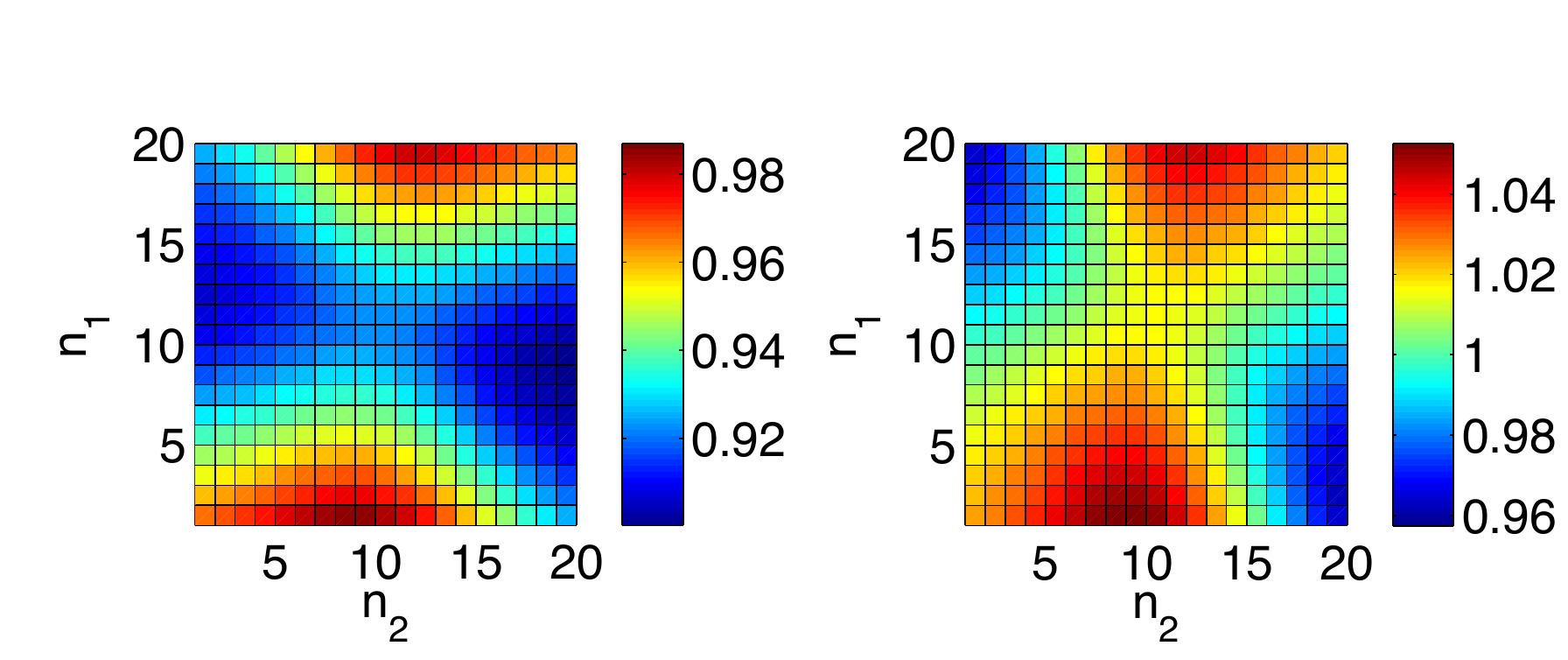}
 \caption{Data  at $T=0.1t$ and for wavevector transfer $(\pi,0)$. In the left plot, we added the three fRG flow results with the right hand side reduced either to cRPA, the particle-particle channel, or the crossed particle-hole channel, divided by 3. In the right plot we only added the separate flows in the particle-particle channel and the crossed particle-hole channel (shown in the left and right plots of Fig. \ref{OFUsc}). These plots should be compared to the full fRG result in the lower right plot of Fig. \ref{OFU10_-50}. The agreement is better for the two-channel sum in the right plot.
 Again, the vertical axis contains the first incoming frequency $\omega_1$, and the horizontal axis the second incoming frequency $\omega_2$. $\omega_3$ is fixed at $-\pi T$. The colorbar illustrates the reduction or enhancement with respect to the bare value $V_{\lambda=0} (\omega_1,\omega_2,\omega_3)=t$.}
\label{OFU23channels}
\end{figure}

Let us now come back to the frequency dependence. In the upper right plots of Figs. \ref{OFU10_100}, \ref{OFU25_100}, and  \ref{OFU10_-50} we show cuts through the frequency dependence of the effective interactions at $\lambda=1$, as function of the frequency transfer between first incoming and first outgoing particle. On a qualitative level, along this cut, the frequency dependence is similar in cRPA and fRG, but with a stronger reduction in cRPA when the filling is such that there is more particle-hole phase space. The difference in the frequency dependence of the effective interactions is best visible in the lower plots of these figures, where we compare $V_{\lambda=1} (\omega_1, \omega_2,\omega_3)$ as function on the incoming frequencies $\omega_1$ and $\omega_2$ with the first outgoing frequency $\omega_3$ fixed to the smallest positive frequency.  While the cRPA results shown in the left lower plots only depend on $\omega_1-\omega_3$ and just show the suppression for small $\omega_1-\omega_3$, the full fRG results are a more complicated function of three frequencies. The main suppression now occurs in the upper left and lower right corners. The line connecting these corners is the line with small or zero total frequency, and for these combinations the repulsive interaction is 'Kanamori'- or 'Anderson-Morel'-screened in the particle-particle channel. The enhancement also seen in the plot can be traced back to the influence of the crossed particle-hole channel that increases the repulsion for small crossed transfer frequency $\omega_2-\omega_3$.  In Fig. \ref{OFUsc} we show the results for the effective conduction-band interactions when only the crossed particle-hole channel or only the particle-particle channel is allowed on the right hand side of the fRG flow equation. These flows are then equivalent to an infinite-order summation in the respective channel. In these plots one can clearly see the enhancement for small transfer $\omega_2-\omega_3$ due to the crossed particle-hole channel and the suppression of interactions with small $\omega_1+\omega_2$ in the particle-particle channel. If one adds these two features at $\lambda=1$, i.e. after doing infinte-order summations in each channel separately, one actually gets near the full fRG results with all channels coupling and active (i.e. in addition also the RPA channel and the vertex corrections), although quantitatively the differences are still clearly visible. Adding as well the summed RPA does not improve the result. The attempts to model the result of the full fRG with coupled channels by the adding single-channel summations are shown in the left plot of Fig. \ref{OFU23channels} for particle-particle ladder sum, crossed particle-hole sum and RPA sum added together, and on the right side of Fig. \ref{OFU23channels} for particle-particle channel and crossed particle-hole channel summed separately and added afterwards. Hence, to get the result quantitatively as correct as possible, one has to do the full flow. 

The data shown in the figures above is for a specific choice of wavevectors $\vec{k}_1$ to $\vec{k}_3$, but other combinations show very similar results. This basically reflects the fact that the wavevector-dependece of the interband and pure high-energy loop diagrams is rather weak due to the energy separation.

Summarizing these results we see that without additional physical ingredients and for onsite bare interactions without wavevector structure, the cRPA does in certain cases present a satisfactory approximation to the effective interactions on an average level, but not regarding the precise frequency structure. In the light of the weak wavevector-dependence of the diagrams just mentioned it is not overly surprising that all one-loop diagrams have a comparable influence. The Wick-ordered fRG scheme allows one to overcome this difficulty and to include all one-loop diagrams, i.e. to do the full parquet summation, in order to obtain a more robust description of the effective interactions. 

Of course, as the one-loop diagrams do not necessarily dominate higher-loop terms for the setup here, even corrections beyond the one-loop level might still have some influence, but to include these or do check their importance might be a very hard task. Furthermore, in our simple model, the quantitative corrections in the effective interactions compared to the bare interactions were not too strong, and hence one might wonder if second order perturbation theory would not be sufficient. We have not explicitly analyzed this question quantitatively, but from the slightly non-linear flows as function of $\lambda$ in the upper left plots of Figs. \ref{OFU10_100}, \ref{OFU25_100}, and  \ref{OFU10_-50} one can observe that there are mild corrections beyond second order. The rationale behind this argument is that if one performs just one  discrete step in the $\lambda$-flow, the corrections will effectively be of second order in the interactions before this step, as  the right hand side of the flow equation is of second order. The non-linearity of the curves shows that the integrated flow of many steps is not just an extrapolation of the first step, but that the second-order corrections of one step change the corrections collected in the next step. This way a series of arbitrarily high order in the bare couplings is generated. If we used stronger bare interactions, the higher-order corrections would be more visible. Note that our goal was to present a flexible mechanism that can go beyond finite-order perturbation theory, and the differences between the different approximations are already visible at this value of the interactions. Therefore  we refrain from discussing for which parameters a second-order approximation might be suitable or not.

\subsection{Effect of wave-vector dependent bare interactions}
As already argued in the introduction, a priori and without additional arguments there is no reason why there should not be significant correction to RPA in the case of wave-vector-independent bare interactions.
We just saw explicitly that that for this case the cRPA, despite capturing important contributions, does not provide the full picture. A main opponent turned out to be the crossed particle-hole channel, which tends to increase the (repulsive) interaction rather than to suppress it. Of course both these particle-hole (PH) channels can be summed up to infinite order separately. The main differences between the two channels is that the RPA channel has an extra fermion-loop minus sign which causes the different sign (suppression/enhancement) in the renormalization due to these two channels. We can also use this simplified approach in order to see what additional effects are caused by a wavevector structure of the bare interactions. In this argument we now drop the frequency structure, that can be added in later, at least for some parts of the outcome.
In order to be precise let us now choose the bare interaction vertex  as
\begin{equation}
\label{ }
V_{\lambda=0} (\vec{k}_1,\vec{k}_2,\vec{k}_3) = v(\vec{k}_1-\vec{k}_3) \, . 
\end{equation}
Here $v(\vec{q})$ can be understood as the Fourier transform of the bare Coulomb interaction $\sim 1/r$, which gives $4\pi e^2 /q^2$ in three dimensions or related expression for layered or lower-dimensional systems, e.g. $2\pi/q$ for stacked perfectly two-dimensional layers and $q\to 0$. The main point is that is interaction is strongly peaked and singular at $q=0$. On the other hand, $\int d^D \, v(\vec{q})$ should remain integrable in the relevant dimension $D$. 

Let us now insert this bare interaction into the RPA and crossed PH channels. The main point is the well-known fact that the different wavevector transfers $\vec{q}$ no not couple in the RPA sum, and for external wavevector transfer $\vec{k}_1- \vec{k}_3=\vec{q}$  every interaction in the sum is $v(\vec{q})$. So we simply get for the cRPA series
\begin{equation}
\label{ }
V_{\mathrm{cRPA}} (\vec{k}_1,\vec{k}_2,\vec{k}_3) = \frac{v(\vec{k}_1-\vec{k}_3)}{1 + 2 v(\vec{k}_1-\vec{k}_3) |P_c(\vec{k}_1-\vec{k}_3)|} \, , 
\end{equation}
with the constrained polarization bubble $P_c(\vec{k}_1-\vec{k}_3)$ that does not contain intra-conduction band screening. This term is real in the limit of zero frequency and we have written the absolute value in order to avoid confusion regarding the sign convention. The factor of two in the denominator is due to the spin sum for $s=1/2$.
As said before, due to the energy separation of the high-energy bands, we expect a rather weak wavevector dependence of this bubble, i.e. we can set $P_c(\vec{k}_1-\vec{k}_3)=\bar P_c$ with some averaged constrained polarization. In contrast with this, in the crossed channel, the wavevector transfer appearing in the bare interactions is $\vec{k}_2-\vec{k}_3 \not= \vec{q}$. The crossed PH sum then gives
\begin{equation}
\label{ }
V_{\mathrm{crPH}} (\vec{k}_1,\vec{k}_2,\vec{k}_3) = \frac{v(\vec{k}_2-\vec{k}_3)}{1  - v(\vec{k}_2-\vec{k}_3) |\bar{P}_c|} \, , 
\end{equation}
Now, there is no factor of two and a minus sign in the denominator, causing an increase of $V_{\mathrm{crPH}} $ with respect to the bare coupling in the repulsive case. Obviously these two types of corrections plus the other PH diagrams and the PP channel compete, unless we focus on a specific wavevector combination, e.g. the direct forward scattering channel with $\vec{k}_1- \vec{k}_3=\vec{q} \to 0$. If now $\vec{k}_2-\vec{k}_3$ is such that the dimensionless $|v(\vec{k}_2-\vec{k}_3) P_c|$ is small, the cRPA sum dominates all other corrections, and the reduction in the cRPA channel can be  much stronger than possible increases in other channels. Then the frequency structure of the effective interactions will be dominated by the RPA channel as well, i.e. the effective interaction will mainly depend on the direct transfer $\omega_1-\omega_3$ and the suppression will be largest when  $\omega_1-\omega_3$ is small.

This way we see that there is a range of external wavevectors where the cRPA sum constitutes the main correction to the bare interaction. If we choose a true long-range Coulomb interaction as bare interaction, there will certainly get into such a situation, if we make $\vec{q}$ only small enough. However, the interesting question is now whether this small $\vec{q}$-regime is significant for the effective local interaction parameters, or whether these are better represented by interactions for less special external wavevectors. If the small-$q$-range is dominant, the next issue is if it is then appropriate to truncate the effective low-energy interaction in its range, i.e. to keep only local and nearest neighbor-terms. The answer to these questions can only be found by more realistic calculations. We hope that the formalism proposed here can help to clarify these issues.

\section{Conclusions and Outlook}
We have presented a functional renormalization (fRG) group framework that allows one to compute the effective interactions of subsets of 'target bands' in multiband models. The subset of target bands is typically at lower excitation energies near the Fermi level, and integrating out the higher-energy bands by the fRG leads to renormalizations of the effective interactions that can have important impact on the low-energy phase diagram. By computing the Wick-ordered correlations rather than one-particle irreducible (1PI) vertices, additional contributions corresponding to the feedback of higher-order vertices in the1PI-scheme are already contained on the four-point truncation level. These contributions represent 'mixed' diagrams with internal lines both in the high-energy sector and in the low-energy bands which would be missed in the usual effective action at the same truncation level.
The RPA particle-hole part of these mixed diagrams is responsible for a screening of the bare interactions by interband particle-hole pairs, and can have a rather strong quantitative effect. These terms were first summed in the so-called constrained RPA scheme (cRPA)\cite{cRPA} that has become a widely used tool for the determination of (local) effective interaction parameters in correlated multiband systems\cite{miyake,sasioglu}. Here, the inclusion of these corrections and the corresponding changes in the interaction parameters can lead to relevant differences of quantitative nature, e.g. of the energy scales for ordering tendencies\cite{uebelacker,smaier}, of of more qualitative character, e.g. by shifting transition lines between different ground states or by opening or closing gap nodes in the case of unconventional superconductors\cite{maier,platt}.

The fRG approach presented here can in principle be reduced to the cRPA content by dropping all diagrams outside the RPA channel. In a simple two-band model with structureless bare interactions we have shown however that the cRPA can result in qualitatively different effective interactions compared to the full fRG with all one-loop diagrams included. In particular, the frequency dependence (on the imaginary axis) of the full fRG effective interaction is much more complex, and the particle-particle channel is clearly visible there as another strong source of a reduction of the effective interactions. Depending on the parameters, the overall suppression can be weaker than that found in cRPA. 
We have argued that for a sharper wavevector-structure of the bare interactions there may well be wavevector combinations where the cRPA is closer to the full results. This required specific care to the long-range or small-$q$-part of the interactions. The question that arises from this study is whether this wavevector regime is somehow representative for the local interaction parameters, and whether it is then justified to ignore longer range effective interactions.

As stated before, the current fRG study with very coarse wavevector resolution is not capable of answering these questions more directly, and we do not attempt to study more realistic situations here. A main obstacle for improving the wavevector resolution is the numerical effort. In our straightforward discretization of the vertex we already kept $N_\omega^3$ frequencies with $N_\omega =20$. For each frequency combination we have $2 \times N_{\vec{k}}^3$  wavevectors for 2 bands and $N_{\vec{k}}=2$. Significant improvements in the wavevector resolution and treating more bands are certainly possible with parallel computing, but in order to resolve the small-wavevector-transfer regime addressed in the previous section one might need additional tricks. An useful approximation may be the channel-decomposition developed by Husemann and Salmhofer\cite{husemann} for the wavevector dependence of the interaction. The same strategy was also used by Karrasch et al. \cite{karrasch} for the frequency dependence in impurity problems. Here the main idea is to write or approximate the full vertex that depends on three variables $p_i =(\omega_i , \vec{k}_i)$ with $i=1,2,3$ as sum of three functions that can depend strongly on either the total incoming $p_1+p_2$ or one of the two transfers $p_1-p_3$ or $p_2-p_3$, but that depend only weakly on the respective other, remnant combinations. Instead of a numerical object of the order of $(N_\omega N_{\vec{k}})^3$ one then deals with a array of order $\nu (N_\omega N_{\vec{k}})$, where $\nu$ is at least $3$ for the three functions just mentioned, or somewhat higher for different form factors that may be used to capture the milder dependence on the remnant variable combinations. Hence, the effort is comparable to that of the cRPA (up to the factor $\nu$) and it should be possible, as a next step, to go significantly beyond the simple toy model studied here and to treat realistic, more complex systems with increased wavevector resolution. We hope that this may help to estimate the usefulness of the scheme presented here in the context of multi-band systems. In any case, our work clarifies the connection between the cRPA scheme and the framework of effective actions in many-particle systems. 

Acknowledgements: We thank Stefan Maier, Stefan Uebelacker, Manfred Salmhofer, Masatoshi Imada and Ryotaro Arita for useful discussions. This work was supported by the DFG priority program SPP1458 on iron arsenides and the DFG research unit FOR 723 on fRG methods.

\end{document}